\documentclass[a4paper,11pt]{article}
\usepackage{pos}
\usepackage{microtype}
\usepackage{orcidlink}
\usepackage{hyperref}
\usepackage[export]{adjustbox}
\usepackage{natbib}
\newcommand{\orcidauthorBENNETT}{0000-0002-1678-6701}
\newcommand{\orcidauthorBRITO}{0000-0002-0568-4507}
\newcommand{\orcidauthorHONG}{0000-0002-3923-4184}
\newcommand{\orcidauthorLEE}{0000-0002-4616-2422}
\newcommand{\orcidauthorLIN}{0000-0003-3743-0840}
\newcommand{\orcidauthorLUCINI}{0000-0001-8974-8266}
\newcommand{\orcidauthorPIAI}{0000-0002-2251-0111} 
\newcommand{\orcidauthorVADACCHINO}{0000-0002-5783-5602}
\newcommand{\orcidauthorZIERLER}{0000-0002-8670-4054}

\title{Progress on meson-glueball mixing in lattice gauge theories coupled to fermions in multiple representations}
\ShortTitle{Meson-glueball mixing in lattice gauge theories with fermions in multiple representations}

\affiliation[a]{ Centre for Mathematical Sciences, University of Plymouth, Plymouth, PL4 8AA, United Kingdom}
\affiliation[b]{ Centre for Quantum Fields and Gravity, Faculty  of Science and Engineering, Swansea University, Singleton Park, SA2 8PP, Swansea, United Kingdom}
\affiliation[c]{ Swansea Academy of Advanced Computing, Swansea University (Bay Campus), Fabian Way, SA1 8EN Swansea, Wales, United Kingdom}
\affiliation[d]{ Department of Physics, Pusan National University, Busan 46241, Korea}
\affiliation[e]{ Extreme Physics Institute, Pusan National University, Busan 46241, Korea}
\affiliation[f]{ Particle Theory and Cosmology Group, Center for Theoretical Physics of the Universe, Institute for Basic Science (IBS), Daejeon, 34126, Korea }
\affiliation[h]{ Institute of Physics, National Yang Ming Chiao Tung University, 1001 Ta-Hsueh Road, Hsinchu 30010, Taiwan}
\affiliation[i]{ Physics Division, National Centre for Theoretical Sciences, Taipei 106319, Taiwan}
\affiliation[j]{ Centre for High Energy Physics, Chung-Yuan Christian University, Chung-Li 32023, Taiwan}
\affiliation[k]{ Department of Mathematics, Faculty of Science and Engineering, Swansea University (Bay Campus), Fabian Way, SA1 8EN Swansea, Wales, United Kingdom}
\affiliation[l]{ Department of Physics, Faculty  of Science and Engineering, Swansea University, Singleton Park, SA2 8PP, Swansea, Wales, UK}
\affiliation[m]{Technical University of Munich, Physics Department, James-Franck-Str.\,1, 85748 Garching, Germany}

\author*[a]{Nuno Brito\,\orcidlink{\orcidauthorBRITO}}
\emailAdd{nuno.rebelobrito@plymouth.ac.uk}
\author[b,c]{Ed Bennett\,\orcidlink{\orcidauthorBENNETT}}
\author[d,e]{Deog~Ki Hong\,\orcidlink{\orcidauthorHONG}}
\author[f]{Jong-Wan Lee\,\orcidlink{\orcidauthorLEE}}
\author[h,i,j]{C.-J. David Lin\,\orcidlink{\orcidauthorLIN}}
\author[k]{Biagio~Lucini\,\orcidlink{\orcidauthorLUCINI}}
\author[b,l]{Maurizio Piai\,\orcidlink{\orcidauthorPIAI}}
\author[a]{Davide Vadacchino\,\orcidlink{\orcidauthorVADACCHINO}}
\author[m,b,l]{Fabian Zierler\,\orcidlink{\orcidauthorZIERLER}}

\onbehalf{on behalf of the TELOS collaboration \vspace{4pt}\\ \centering 
\href{https://telos-collaboration.github.io}{ \includegraphics[height=1cm]{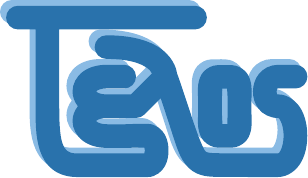} } }

\abstract{We present preliminary results from our numerical lattice investigation of the low-lying flavour-singlet bound states in the ${\rm Sp}(4)$ gauge theory coupled to fermions transforming in the fundamental and two-index antisymmetric representations. We use the Wilson gauge action and Wilson fermions. We employ the generalized eigenvalue problem (GEVP) method with a basis of smeared glueball and meson interpolating operators. We analyse two high-statistics ensembles with the same lattice coupling and different fermion mass. We study both the scalar and pseudoscalar states. We observe the first evidence for meson–glueball mixing in this theory. We improve upon previous determinations of the energy eigenstates and mixing angles in the flavour-singlet pseudoscalar sector.}

\FullConference{The 43rd International Symposium on Lattice Field Theory (LATTICE2026)\\
 26 July - 1 August 2026, 
Maryland, USA}

\begin{document}
\begin{flushright}
\textsc{TUM-EFT 221/26}
\vspace{-0.75cm} 
\end{flushright}

\newcommand{\conn}[1]{\mathcal{C}_{#1}}   
\newcommand{\disc}[1]{\mathcal{D}_{#1}}   
\newcommand{\dgscale}{0.8}               
\newcommand{\dg}[1]{\vcenter{\hbox{%
  \includegraphics[page=#1, scale=\dgscale]{figures/other/correlator_matrix_diagrams.pdf}}}}
\newcommand{\Nr}[1]{N_{\mathrm{#1}}}       

\maketitle

\section{Introduction}

Prominent candidates for extensions of the Standard Model of particle physics are based on the ${\rm Sp}(4)$ gauge theory coupled to an admixture of $N_{\rm f}=2$ and $N_{\rm as}=3$ Dirac fermions,  $Q^I$ and $\Psi^k$,  transforming, respectively,  in the fundamental and antisymmetric representation~\cite{Barnard:2013zea,Ferretti:2013kya}. 
The enhanced, global ${\rm U}(4)\times {\rm U}(6)$ symmetry of the Lagrangian density is broken spontaneously and explicitly, to the ${\rm Sp}(4)\times {\rm SO}(6)$ subset, by the diagonal mass terms, and the  aligned condensates. This is the minimal theory amenable to lattice numerical studies that provides a simultaneous realisation of the paradigms of Composite Higgs Models (CHMs)~\cite{Kaplan:1983fs, Georgi:1984af, Dugan:1984hq}, and top partial compositeness (TPC)~\cite{Kaplan:1991dc}. One linear combination of the two $U(1)$ factors is anomaly free, adding phenomenological interest~\cite{Belyaev:2015hgo, DeGrand:2016pgq, Cacciapaglia:2019bqz}. 

The enhanced global symmetries and associated model-building opportunities motivate the dedicated programme of lattice studies by the TELOS collaboration~\cite{Bennett:2017kga}.  The properties of flavoured mesons and baryons with (heavy) dynamical fermions have been investigated~\cite{Bennett:2022yfa,Bennett:2024cqv,TELOS:2025ash}, as well as the closely related dynamical theories with either fundamental~\cite{Bennett:2019jzz,TELOS:2026alk,Bennett:2026cev} or antisymmetric~\cite{Bennett:2023gbe,Bennett:2024tex} matter fields, and the quenched theory~\cite{Bennett:2019cxd,Bennett:2023mhh,Bennett:2023qwx}.
Meson singlets~\cite{Bennett:2023rsl,Bennett:2024wda} and the glueballs~\cite{Bennett:2020qtj} have been so far studied separately. We present here the first preliminary measurements of the spin-0, flavour-singlet spectroscopy that combine meson and glueball operators of this theory.
We set the scale as in Ref.~\cite{Bennett:2022ftz}, and provide the first estimates of mixing between mesons and glueballs.

\section{Lattice observables}

We discretize the ${\rm Sp}(4)$ gauge theory with the Wilson plaquette action, supplemented by Wilson-Dirac fermions for the matter fields. Ensembles are generated using the Hybrid Monte Carlo (HMC) algorithm implemented in the Grid software environment~\cite{Boyle:2015tjk,Boyle:2016lbp,Yamaguchi:2022feu,Grid-repo}, supplemented by the TELOS implementation of the ${\rm Sp}(4)$ group~\cite{Bennett:2023gbe} and run on GPU architecture machines. We have increased the statistics of two existing ensembles~\cite{Bennett:2024wda}, that have common lattice coupling, $\beta = 6.5$, and bare antisymmetric fermion mass, $am_{0}^{\rm as}=-1.01$, but different temporal lattice extensions, $N_{t}=64,\,96$, and fundamental fermion mass, $am_{0}^{\rm f}=-0.7,\,-0.71$---see Table~\ref{tab:ensembles}.
For measurements we use HiRep~\cite{HiRepSUN,HiRepSpN} on CPU machines. Meson and glueball operators of interest are schematically written as 
\begin{align} 
    \label{eq:interpolating_operators}
    O_{\rm (R)}^{\Gamma}(x)  &= \frac{1}{\sqrt{N_{\rm (R)}}} \sum_{I=1}^{N_{\rm (R)}} \bar  \Phi^I(x) \Gamma \Phi^I(x)\,,\, \quad O_{\rm G}^{A_{1}}[L](x) = \sum_{r\in O_{\rm h}} {\rm Tr}~r[L(x)]\,,
\end{align}
where $\Phi = Q,\,\Psi$ are the spinors, $(R) = ({\rm f}, {\rm as})$ identifies the representation of ${\rm Sp}(4)$, and $\Gamma = \mathbb{I}, \gamma_5$ denotes the Dirac matrices---notation and details can be found in Ref.~\cite{Bennett:2024wda}. The glueball operator, $O_{\rm G}^{A_{1}}[L](x)$, is built as a linear combination of the rotations, $r$, applied on a prototype Wilson loop, $L$,  projected onto the irreducible representation, $A_{1}$, of the octahedral group $O_{h}$, with zero momentum. 
\begin{table}[h]
\centering
\caption{Characterisation of the ensembles used in this work. Beside the fundamental fermion mass, $am_0^{\rm f}$, and the temporal and spatial extent of the lattice, $N_t$ and $N_s$, we report the number, $N_{\rm therm}$, of HMC thermalization updates,  and of skipped  updates between measurements, $n_{\rm skip}$. The average plaquette, $\langle P\rangle$, is reported together with its integrated autocorrelation time, $\tau_{\rm int}^P$,  the  Wilson-flow  scale, $w_{0}/a$, the average topological charge, $\bar{Q}$, and its  integrated autocorrelation time, $\tau^{Q}_{\rm int}$. The lattice coupling is  $\beta = 6.5$ and the ${\rm as}$-type mass $am_{0}^{\rm as} = -1.01$. 
    \label{tab:ensembles}
    }
    {\setlength{\tabcolsep}{3pt}
    \begin{tabular}{|c|c|c|c|c|c|c|c|c|c|c|c|c|c|}
    \hline \hline
    Label & $am_0^{\rm f}$ & $N_t$ & $N_s$ & $N_{\rm therm}$ & $n_{\rm skip}$ & $N_{\rm conf}$ & $\langle P \rangle$ & $\tau_{\rm int}^P$ & $w_0 / a$ & $\bar{Q}$ & $\tau_{\rm int}^Q$ \\ 
    \hline
        M3 & -0.71 & 96 & 20 & 4000 & 26 & 4507 & 0.585156(13) & 2.0(4) & 2.5170(40)  & -0.60(19) & 6.4(3.3) \\ 
        M4 & -0.70 & 64 & 20 & 1000 & 20 & 5683 & 0.584228(12) & 2.4(9) & 2.3557(31)  & -0.31(19) & 10.6(4.8) \\ 
    \hline \hline
    \end{tabular}
    }
\end{table}
\newpage The two-point correlation functions of interest have the following  matrix form
\begin{eqnarray}
    C(t-t') = {\setlength{\arraycolsep}{5pt}
  \begin{pmatrix}
    \disc{GG} & \dfrac{1}{\sqrt{\Nr{f}}}\,\disc{GQ} & \dfrac{1}{\sqrt{\Nr{as}}}\,\disc{G\Psi} \\[8pt]
     & \conn{Q} - \dfrac{1}{\Nr{f}}\,\disc{QQ} & \dfrac{1}{\sqrt{\Nr{f}\Nr{as}}}\,\disc{Q\Psi} \\[8pt]
     & & \conn{\Psi} - \dfrac{1}{\Nr{as}}\,\disc{\Psi\Psi}
  \end{pmatrix}}\,.
  \end{eqnarray}
  The entries are
expressed in terms of the following diagrams:
  \begin{eqnarray}
  \begin{array}{@{}r@{\;=\;}l@{\quad}r@{\;=\;}l@{\quad}r@{\;=\;}l@{\quad}r@{\;=\;}l@{}}
    \disc{GG} & \dg{1} & \disc{GQ}  & \dg{2} & \disc{G\Psi} & \dg{3} & \disc{Q\Psi}    & \dg{6} \\[6pt]
    \conn{Q}  & \dg{4} & \disc{QQ}  & \dg{5} & \conn{\Psi}  & \dg{7} & \disc{\Psi\Psi} & \dg{8}
  \end{array}
  \notag
\end{eqnarray}
where the solid (dashed) line denotes the contraction of two fundamental (antisymmetric) fermion fields and the blue lines denote a glueball operator. More details will be presented in our future publication, including our implementation of APE and Wuppertal smearing, as well as partial pruning~\cite{Urrea-Nino:2025ztj,juan_mixing_swave},
which we use to optimise the signal and reduce numerical instabilities, respectively. The energy levels of the theory are then extracted via a variational procedure based on a Generalized Eigenvalue Problem (GEVP), which we define as follows:
\begin{align} \label{eq:GEVP}
    C(t_{1})v(t_{0},t_{1}) = \lambda(t_{0},t_{1})C(t_{0})v(t_{0},t_{1})\,.
\end{align}
The eigenvectors, $v(t_{0}, t_{1})$, are then used to diagonalise the correlation matrix, after which we fit the diagonal elements to the following exponential form,
\begin{equation}
\label{eq:diagonalized_correlators}
    D(t) \equiv v^{T}(t_{0},t_{1})C(t)v(t_{0},t_{1}) \xrightarrow[\rm fit]{} A\left(e^{-E_{n}t} + e^{-E_{n}(a N_{t} - t)}\right)\,.
\end{equation}
The operator $\mathcal{O}^{\dagger}_{n} \equiv \sum_{i} v^{n}_i(t_{0},t_{1})O_{i}^{\dagger}$ can be interpreted as our best estimate of the operator that sources the state with energy $E_{n}$. In practice, we found best numerical stability when setting $t_{1}= t_{0}+a$ always, and $t_{0} = 0$ when possible, with the exception that, in the scalar case, whenever mesonic operators are included, we set $t_{0} = a$. 

The final estimates for the energy levels are obtained from a model-weighted average based on the Akaike Information Criterion (AIC)~\cite{Akaike:1998zah, Jay:2020jkz}, combining values of $E_{n}$ extracted from fitting $D(t)$ over different time intervals. Having defined  the effective mass as,
\begin{equation}
\label{eq:effective_mass}
    am_{\rm eff}(t) = \operatorname{arccosh}\left(\frac{D(t+a) - D(t-a) }{2D(t)}\right)\,,
\end{equation}
we perform a visual check by plotting the effective mass together with the AIC results. 

In analogy to QCD~\cite{Michael:2013gka,Ottnad:2025zxq}, we write the two lightest flavour-singlet pseudoscalars, $\eta_{l}'$ and $\eta_{h}'$, as mixtures of single representation, \textit{flavour} states $\eta_{\rm f}'$ and $\eta_{\rm as}'$, 
\begin{equation}
\label{eq:mixing_angles_matrix}
    \begin{pmatrix}
        | \eta'_{l}\rangle \\
        | \eta'_{h} \rangle 
    \end{pmatrix} 
    =
    \begin{pmatrix}
    A_{\eta'_l }\,{\rm \cos}~\phi_{l} &  A_{\eta'_l }\,{\rm \sin}~\phi_{l} \\
    -A_{\eta'_h }\,{\rm \sin}~\phi_{h} & A_{\eta'_h }\,{\rm \cos}~\phi_{h}
    \end{pmatrix}
    \cdot
    \begin{pmatrix}
    | \eta'_{\rm f}\rangle \\
    | \eta'_{\rm as} \rangle
    \end{pmatrix}\,,
\end{equation}
where  $A_{\eta'_l}, A_{\eta'_h }$ are two constants. $\phi_l$ and $\phi_h$ are the two mixing angles, that can be determined from
\begin{equation}
    \label{eq:effective_mixing_angles_explicit}
        \tan(\phi_{l}) = \frac{\langle0| O^{\gamma_5}_{\rm as} |\eta'_{l}\rangle}{\langle0| O^{\gamma_5}_{\rm f} |\eta'_{l}\rangle}\,,
        \qquad
        -\tan(\phi_{h}) = \frac{\langle0| O^{\gamma_5}_{\rm f} |\eta'_{h}\rangle}{\langle0| O^{\gamma_5}_{\rm as} |\eta'_{h}\rangle}\,.
 \end{equation}
 We find it useful to define also the combination~\cite{Michael:2013gka,Ottnad:2025zxq},
 \begin{equation}
     \label{eq:effective_mixing_angles_explicit_combination}
        -\tan(\phi)^{2} \equiv \frac{\langle0| O^{\gamma_5}_{\rm as} |\eta'_{l}\rangle}{\langle0| O^{\gamma_5}_{\rm f} |\eta'_{l}\rangle} \cdot \frac{\langle0| O^{\gamma_5}_{\rm f} |\eta'_{h}\rangle}{\langle0| O^{\gamma_5}_{\rm as} |\eta'_{h}\rangle}\,.
\end{equation}
The angles defined in Eqs.~\eqref{eq:effective_mixing_angles_explicit},\,\eqref{eq:effective_mixing_angles_explicit_combination} are estimated using the overlap functions of unsmeared meson operators, $ O_{\rm f}^{\gamma_5}$ and $O_{\rm as}^{\gamma_5}$. 

\begin{figure}[t]
    \centering
    \includegraphics{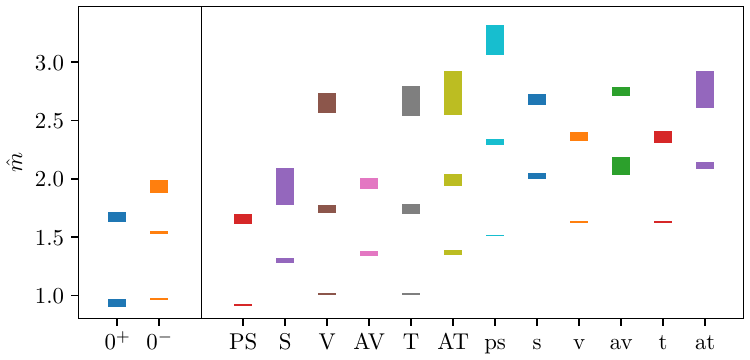}
    \hfill 
    \caption{Preliminary numerical results for the masses of mesons in the M3 ensemble of Tab.~\ref{tab:ensembles}, expressed in units of $w_0$. The flavour singlet pseudoscalar and scalar, denoted $0^{-}$ and $0^{+}$, respectively, are new in this work, while the flavoured states are taken from Ref.~\cite{Bennett:2024cqv}. Uppercase (lowercase) labels denote fundamental (antisymmetric) constituents. Boxes represent the energy levels and their statistical uncertainties. \label{fig:meson_spectrum_plot}}
\end{figure}

The treatment of meson-glueball mixing follows the same principles, and will be discussed in details in an upcoming publication. We define an effective mixing angle, $\theta$, via the relation
\begin{equation}
\label{eq:meson_glueball_angle}
\cos\left(2\theta\right) \equiv
  \frac{  E_{0}^{(G)} - E_{0}^{(M)}}{E_{1} - E_{0}} \,\simeq\, 
   \frac{  E_{0}^{(G)} - E_{0}^{(M)}}{E_{1}^{(G)} - E_{0}^{(G)}} \,,
\end{equation}
where $E_{0,1}$ are the ground and first excited state energies computed with the full basis of operators, including both mesons and glueballs, 
$E_{0,1}^{(G)}$ are the same energies, but computed using only glueball operators, and $E_{0}^{(M)}$ is the ground state energy obtained with a basis that includes meson operators built with either the fundamental or the antisymmetric fermions. In some cases, the energies $E_{0,1}^{(G)}$ are better determined numerically in comparison to $E_{0,1}$ and the approximate relation in Eq.~\eqref{eq:meson_glueball_angle} holds for the parameter space explored.

\section{Preliminary numerical results}

\begin{figure}[t]
    \centering
    \includegraphics{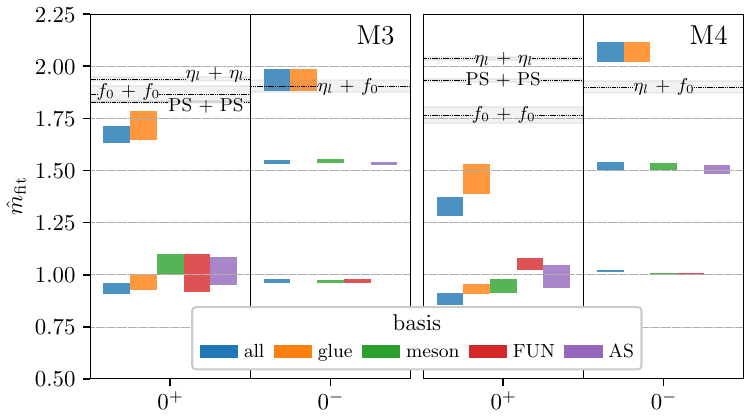}
    \caption{Flavour-singlet scalar ($0^+$) and pseudoscalar ($0^-$) spectra in the  $\mathrm{Sp}(4)$, $N_{\rm f}=2$ and $N_{\rm as}=3$ theory, extracted via GEVP analysis of the two-point correlation functions, for ensembles M3 and M4 in Table~\ref{tab:ensembles}. Different colours denote different choices of GEVP operator basis: all (blue), glueball (orange), meson (green), fundamental (red) and antisymmetric (purple) fermions, respectively. Two-particle thresholds, computed with the full GEVP basis, are shown for comparison.  The lightest $0^-$and $0^+$ states are denoted as $\eta'_{l}$ and $f_0$.}
    \label{fig:basis_comparison_plot}
\end{figure}

In Fig.~\ref{fig:meson_spectrum_plot}, we display preliminary results for the masses, expressed in units of the Wilson flow scale, $\hat{m}\equiv m w_0$, of the lightest, flavour-singlet, scalar and pseudoscalar states, labelled $0^{+}$ and $0^{-}$ for the M3 ensemble. The values were obtained using the complete operator basis in the GEVP analysis, including glueball and meson operators. For illustration purposes, we show also the flavoured mesons, taken from  Ref.~\cite{Bennett:2024cqv}. Note that, in both ensembles, dynamical fermions are comparatively heavy. We find that the masses for the flavour-singlet scalar and pseudoscalar ground states are close to those appearing in the flavoured pseudoscalar (PS), vector (V) and tensor (T) channels and are notably lighter than the rest of the spectrum. 

Figure~\ref{fig:basis_comparison_plot} shows the flavour-singlet energy levels obtained with the full variational basis in the GEVP analysis, compared with the results of measurements obtained by restricting the basis, as indicated in the legend. We also display two-particle thresholds for some relevant channels, estimated using the results from the full GEVP analysis. We denote with $f_{0}$ the lightest $0^+$ singlet state. The value for $\hat{m}_{\rm PS}$ is taken from Ref.~\cite{Bennett:2024cqv}.

In both ensembles, the $0^{-}$  energy levels obtained using different choices of  variational basis in the GEVP analysis  are compatible with each other, when available.  Besides the ground state, $\eta'_{l}$, and first excited state, $\eta'_{h}$, we find indication of a  heavier state with the introduction of glueball operators. This signal is close to the two particle threshold in the M3 ensemble and above it in the M4 ensemble. In the $0^{+}$ channel, we have clear evidence of the ground state, which we call $f_0$, in all analyses, and in the presence of the glueball operators in GEVP we find a signal of a first excited state.   Intriguingly, as we extend the GEVP operator basis the mass decreases, suggesting that  operator mixing is present in the ground state.  The excited state appears below threshold, and its mass decreases in going from M3 to M4.

\begin{figure}[t]
    \centering
    \includegraphics{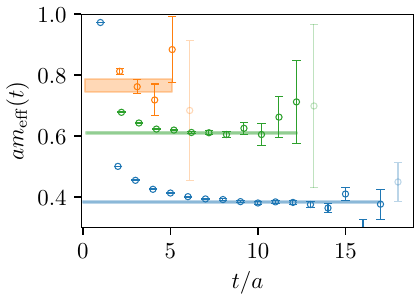}
    \includegraphics{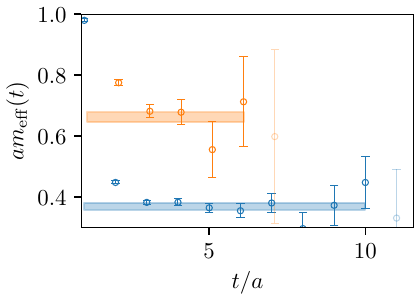}
    \caption{Effective mass, $am_{\rm eff}(t)$, in the $0^-$ (left) and $0^+$ (right) channel, as a function of the temporal separation, $t$, in the M3 ensemble. The lowest non-degenerate solutions of the GEVP analysis are represented by distinct coloured plateaus. Faded data points correspond to measurements with noise-to-signal ratio of $D(t)$ exceeding $10\%$, that are excluded from the analysis. The shaded bands indicate the results obtained from the model-averaging fit procedure, based on single periodic-exponential fits performed over different temporal fitting ranges. The vertical extent of each shaded band corresponds to the extracted energy, $E_{n}$, and its associated uncertainty, while the horizontal extent denotes the range of temporal intervals used in the determination of the fit windows entering the model-averaging procedure. \label{fig:meff}}
\end{figure} 

In Fig.~\ref{fig:meff}, we plot the effective masses, $am_{\rm eff}(t)$, in the M3 ensemble for the $0^-$ and $0^+$ channels. The signal for the ground state and first excited state are clear. The second excited state in the pseudoscalar channel exhibits a short plateau at early Euclidean time. Establishing whether this signal is a truly independent new state requires further  investigation. 

\begin{figure}[t]
    \centering
    \includegraphics{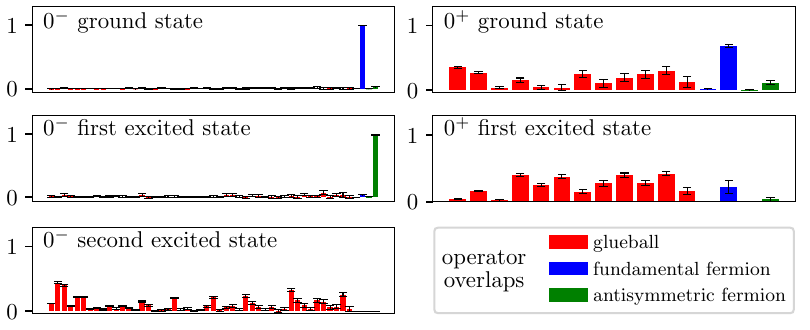}
    \caption{Overlap factors extracted from the GEVP eigenvectors, $v(t_0,t_1)$, for the M3 ensemble. The panels in the left (right) column show the overlaps, as a histogram plot, for the  $0^-$ and $0^+$ channels.  Each row corresponds to a different energy level (ground state, first excited state, second excited state). Each bar corresponds to one operator, coloured by type: glueball (red), fundamental meson (blue) and antisymmetric meson (green). Bar heights give central values, and error bars give the standard deviation of the overlap distribution. The number of bins corresponds to the number of operators left after applying partial pruning.
    \label{fig:overlaps}} 
\end{figure}

In Fig.~\ref{fig:overlaps}, we present the overlap factors, extracted from the GEVP analysis, for each flavour-singlet state. In the $0^-$ channel, the ground state, $\eta'_{l}$, and the first excited state, $\eta'_{h}$, are mostly sourced by a single type of meson operator, built of fundamental and antisymmetric fermions, respectively. The second excited state appears to be sourced mostly by glueball operators. This further supports the interpretation that mixing effects are small in the $0^-$ channel. This should be contrasted with the $0^+$ channel. Neither meson nor glueball operators dominate the overlaps. This indicates mixing effects between glueball and meson singlets in this theory is significant, much like in the flavour-singlet sector of QCD \cite{UKQCD:2006singlet0pp, Brett:2019scalarglueball, Gui:2026adh,juan_mixing_swave,juan_mixing_charmonium_glueball}. 

\begin{figure}[t]
\centering
\includegraphics{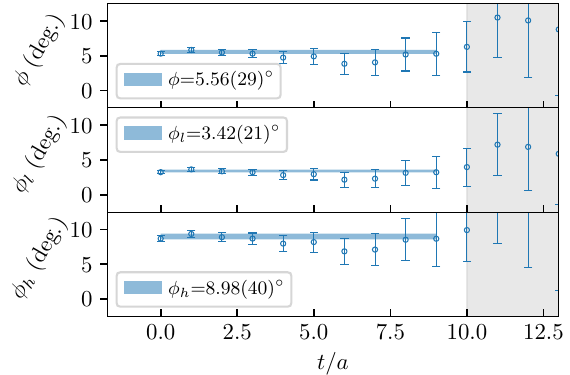}
\includegraphics{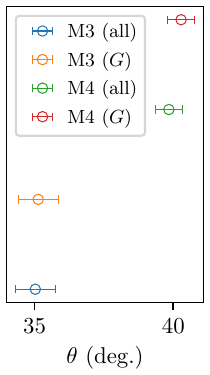}
\caption{Left panel: $\eta'_{l}-\eta'_{h}$ mixing angles measured in the M3 ensemble, using a single-angle parametrization, $\phi$ (top panel), or two-angle parametrization, $\phi_{l}-\phi_{h}$ (middle and bottom panel). The gray shaded area represents the region where the relative error of the ground state effective mass is over 10\%. Right panel: meson-glueball mixing angle, $\theta$,  extracted, for  ensembles M3 and M4, from energy measurements using the full GEVP operator basis (all), restricted to glueball operators ($G$), see  Eq.~\eqref{eq:meson_glueball_angle}. All angles are expressed in degrees. \label{fig:mixing_angles}}
\end{figure}

In Fig.~\ref{fig:mixing_angles}, we display our measurements of the mixing angles, both in the $0^-$ (left) and $0^+$ (right) channels. We can extract the angles $\phi_{l}$, $\phi_{h}$, and $\phi$ in the $0^-$ case, obtaining numerically small but non-zero values, signalling small mixing effects, as expected. The values for the three angles are not compatible with one another, which is also seen in QCD~\cite{Michael:2013gka}, indicating that the single-angle parametrisation may not be the optimal choice. In the $0^+$ channel, the meson-glueball mixing angle turns out to be large, in both ensembles, supporting the presence of large mixing effects. The value of the mixing angles are different for the two ensembles explored.  A similar calculation in one flavour lattice QCD also found large mixing angles, see Ref.~\cite{Gui:2026adh} and references therein. 

\acknowledgments
N.~B is supported by the STFC Consolidated Grant No.~ST/X508676/1. The work of B.~L., E.~B., F.~Z., and M.~P. is supported by the STFC Consolidated Grant No.~ST/X000648/1. B.~L. is supported in part by the STFC Consolidated Grant No.~ST/X00063X/1. E.~B. and M.~P. are supported by STFC [UKRI4243]. D.~K.~H. is supported by Basic Science Research Program through NRF funded by the Ministry of Education (NRF-2017R1D1A1B06033701). The work of D.~V. is supported by the STFC Consolidated Grant No.~ST/X000680/1. C.-J.~D.~L. acknowledges support from NSTC Taiwan through grant number 112-2112-M-A49-021-MY3. C.-J.~D.~L. is also supported by the Taiwanese MoST Grant No.~109-2112-M-009-006-MY3. C.-J.~D.~L. is also supported by Grants No.~112-2639-M-002-006-ASP and No.~113-2119-M-007-013-. J.-W.~L. is supported by IBS under the project code, IBS-R018-D1. The work of E.~B. is supported by the STFC Research Software Engineering Fellowship EP/V052489/1 and has been supported in part by UKRI via the Computational Science Centre for Research Communities (CoSeC), under Grant No.~UKRI497. F.~Z. is supported by the Advanced ERC grant ERC-2023-ADG-Project EFT-XYZ.

{\textbf{High performance computing}}---This work used the DiRAC Data Intensive service (DIaL3) at the University of Leicester, Data Intensive service (CSD3) at the University of Cambridge, and Extreme Scaling service (Tursa) at the University of Edinburgh, managed by the University of Leicester Research Computing Service, University of Cambridge University Information Services, and EPCC respectively on behalf of the STFC DiRAC HPC Facility (www.dirac.ac.uk). The DiRAC service at Leicester, the DiRAC component of CSD3 at Cambridge, and the DiRAC service at Edinburgh were funded by BEIS, UKRI and STFC capital funding and STFC operations grants. DiRAC is part of the UKRI Digital Research Infrastructure. Numerical computations were performed using Supercomputing Wales SUNBIRD at Swansea University. This work was partially carried out using the computational facilities of the \textit{Lovelace} High Performance Computing Centre, University of Plymouth. Supercomputing Wales is supported by the European Regional Development Fund via Welsh Government.

{\bf Research Data Access Statement}---These results are preliminary. Further analysis and the data generated for this manuscript will be released together with an upcoming publication.

{\bf Open Access Statement}---For the purpose of open access, the authors have applied a Creative Commons Attribution (CC BY) license to any Author Accepted Manuscript version arising.

\bibliographystyle{jhep-notitle}
\bibliography{refs}

@article{Boyle:2015tjk,
    author = "Boyle, Peter and others",
    title = "{Grid: A next generation data parallel C++ QCD library}",
    eprint = "1512.03487",
    archivePrefix = "arXiv",
    primaryClass = "hep-lat",
    doi = "10.22323/1.251.0023",
    journal = "PoS",
    volume = "LATTICE2015",
    pages = "023",
    year = "2016"
}

@article{	  Boyle:2016lbp,
  author	= "Boyle, Peter A. and others",
  title		= "{Grid: A next generation data parallel C++ QCD library}",
  doi		= "10.22323/1.251.0023",
  journal	= "PoS",
  volume	= "LATTICE2015",
  pages		= "023",
  year		= "2016"
}

@article{	  Yamaguchi:2022feu,
  author	= "Yamaguchi, Azusa and others",
  title		= "{Grid: OneCode and FourAPIs}",
  eprint	= "2203.06777",
  archiveprefix	= "arXiv",
  primaryclass	= "hep-lat",
  doi		= "10.22323/1.396.0035",
  journal	= "PoS",
  volume	= "LATTICE2021",
  pages		= "035",
  year		= "2022"
}

@misc{Grid-repo,
  author = {{Grid contributors}},
  title = {Grid: Data parallel {C++} mathematical object library},
  howpublished = {\url{https://github.com/paboyle/Grid}}
}

@article{Ferretti:2013kya,
    author = "Ferretti, Gabriele and Karateev, Denis",
    title = "{Fermionic UV completions of Composite Higgs models}",
    eprint = "1312.5330",
    archivePrefix = "arXiv",
    primaryClass = "hep-ph",
    doi = "10.1007/JHEP03(2014)077",
    journal = "JHEP",
    volume = "03",
    pages = "077",
    year = "2014"
}

@article{Barnard:2013zea,
    author = "Barnard, James and others",
    title = "{UV descriptions of composite Higgs models without elementary scalars}",
    eprint = "1311.6562",
    archivePrefix = "arXiv",
    primaryClass = "hep-ph",
    doi = "10.1007/JHEP02(2014)002",
    journal = "JHEP",
    volume = "02",
    pages = "002",
    year = "2014"
}

@article{Cacciapaglia:2019bqz,
    author = "Cacciapaglia, Giacomo and others",
    title = "{Light scalars in composite Higgs models}",
    eprint = "1902.06890",
    archivePrefix = "arXiv",
    primaryClass = "hep-ph",
    reportNumber = "CTPU-PTC-18-39, LU TP 18-41",
    doi = "10.3389/fphy.2019.00022",
    journal = "Front. in Phys.",
    volume = "7",
    pages = "22",
    year = "2019"
}

@article{Bennett:2026cev,
    author = "Bennett, Ed and others",
    title = "{Symplectic lattice gauge theories in the Grid framework: domain wall fermions and continuum extrapolations}",
    eprint = "2609.19930",
    archivePrefix = "arXiv",
    primaryClass = "hep-lat",
    reportNumber = "CTPU-PTC-26-20, PNUTP-26/A02",
    month = "9",
    year = "2026"
}

@article{TELOS:2026alk,
    author = "Bennett, Ed and others",
    collaboration = "TELOS",
    title = "{Resonant scattering in two-flavored Sp(4) lattice gauge theories}",
    eprint = "2606.14546",
    archivePrefix = "arXiv",
    primaryClass = "hep-lat",
    reportNumber = "CTPU-PTC-26-13, TUM-EFT 218/26",
    month = "6",
    year = "2026"
}

@article{TELOS:2025ash,
    author = "Bennett, Ed and others",
    collaboration = "TELOS",
    title = "{Chimera baryons and mesons on the lattice: A spectral density analysis}",
    eprint = "2506.19804",
    archivePrefix = "arXiv",
    primaryClass = "hep-lat",
    reportNumber = "UTCCS-P168, UTHEP-807, PNUTP-25/A04, CTPU-PTC-25-24",
    doi = "10.1103/qkqx-xxb2",
    journal = "Phys. Rev. D",
    volume = "112",
    number = "7",
    pages = "074515",
    year = "2025"
}

@article{Bennett:2024tex,
    author ="Bennett, Ed and others",
    title = "{Meson spectroscopy in the Sp(4) gauge theory with three antisymmetric fermions}",
    eprint = "2412.01170",
    archivePrefix = "arXiv",
    primaryClass = "hep-lat",
    reportNumber = "CTPU-PTC-24-32, UTHEP-794, UTCCS-P-160",
    doi = "10.1103/PhysRevD.111.074511",
    journal = "Phys. Rev. D",
    volume = "111",
    number = "7",
    pages = "074511",
    year = "2025"
}

@article{Bennett:2024wda,
    author = "Bennett, Ed and others",
    title = "{Mixing between flavor singlets in lattice gauge theories coupled to matter fields in multiple representations}",
    eprint = "2405.05765",
    archivePrefix = "arXiv",
    primaryClass = "hep-lat",
    reportNumber = "CTPU-PTC-24-12, PNUTP-24/A03",
    doi = "10.1103/PhysRevD.110.074504",
    journal = "Phys. Rev. D",
    volume = "110",
    number = "7",
    pages = "074504",
    year = "2024"
}

@article{Bennett:2023qwx,
    author = "Bennett, Ed and others",
    title = "{Spectrum of mesons in quenched Sp(2N) gauge theories}",
    eprint = "2312.08465",
    archivePrefix = "arXiv",
    primaryClass = "hep-lat",
    reportNumber = "CTPU-PTC-23-50",
    doi = "10.1103/PhysRevD.109.094517",
    journal = "Phys. Rev. D",
    volume = "109",
    number = "9",
    pages = "094517",
    year = "2024"
}

@article{Bennett:2023gbe,
    author = "Bennett, Ed and others",
    title = "{Symplectic lattice gauge theories in the grid framework: Approaching the conformal window}",
    eprint = "2306.11649",
    archivePrefix = "arXiv",
    primaryClass = "hep-lat",
    reportNumber = "PNUTP-23/A03, CTPU-PTC-23-26",
    doi = "10.1103/PhysRevD.108.094508",
    journal = "Phys. Rev. D",
    volume = "108",
    number = "9",
    pages = "094508",
    year = "2023"
}

@article{Bennett:2023mhh,
    author = "Bennett, Ed and others",
    title = "{Lattice investigations of the chimera baryon spectrum in the Sp(4) gauge theory}",
    eprint = "2311.14663",
    archivePrefix = "arXiv",
    primaryClass = "hep-lat",
    doi = "10.1103/PhysRevD.109.094512",
    journal = "Phys. Rev. D",
    volume = "109",
    number = "9",
    pages = "094512",
    year = "2024"
}

@article{Bennett:2023rsl,
    author = "Bennett, Ed and others",
    title = "{Singlets in gauge theories with fundamental matter}",
    eprint = "2304.07191",
    archivePrefix = "arXiv",
    primaryClass = "hep-lat",
    reportNumber = "CTPU-PTC-23-11",
    doi = "10.1103/PhysRevD.109.034504",
    journal = "Phys. Rev. D",
    volume = "109",
    number = "3",
    pages = "034504",
    year = "2024"
}

@article{Bennett:2022ftz,
    author = "Bennett, Ed and others",
    title = "{Sp(2N) Yang-Mills theories on the lattice: Scale setting and topology}",
    eprint = "2205.09364",
    archivePrefix = "arXiv",
    primaryClass = "hep-lat",
    reportNumber = "PNUTP-22/A03",
    doi = "10.1103/PhysRevD.106.094503",
    journal = "Phys. Rev. D",
    volume = "106",
    number = "9",
    pages = "094503",
    year = "2022"
}

@article{Bennett:2022yfa,
    author = "Bennett, Ed and others",
    title = "{Lattice studies of the Sp(4) gauge theory with two fundamental and three antisymmetric Dirac fermions}",
    eprint = "2202.05516",
    archivePrefix = "arXiv",
    primaryClass = "hep-lat",
    reportNumber = "PNUTP-22/A01",
    doi = "10.1103/PhysRevD.106.014501",
    journal = "Phys. Rev. D",
    volume = "106",
    number = "1",
    pages = "014501",
    year = "2022"
}

@article{Bennett:2020qtj,
    author = "Bennett, Ed and others",
    title = "{Glueballs and strings in $Sp(2N)$ Yang-Mills theories}",
    eprint = "2010.15781",
    archivePrefix = "arXiv",
    primaryClass = "hep-lat",
    doi = "10.1103/PhysRevD.103.054509",
    journal = "Phys. Rev. D",
    volume = "103",
    number = "5",
    pages = "054509",
    year = "2021"
}

@article{Bennett:2019cxd,
    author = "Bennett, Ed and others",
    title = "{$Sp(4)$ gauge theories on the lattice: quenched fundamental and antisymmetric fermions}",
    eprint = "1912.06505",
    archivePrefix = "arXiv",
    primaryClass = "hep-lat",
    reportNumber = "PNUTP-19/A03",
    doi = "10.1103/PhysRevD.101.074516",
    journal = "Phys. Rev. D",
    volume = "101",
    number = "7",
    pages = "074516",
    year = "2020"
}

@article{Bennett:2019jzz,
    author = "Bennett, Ed and others",
    title = "{Sp(4) gauge theories on the lattice: $N_f=2$ dynamical fundamental fermions}",
    eprint = "1909.12662",
    archivePrefix = "arXiv",
    primaryClass = "hep-lat",
    reportNumber = "PNUTP-19/A01",
    doi = "10.1007/JHEP12(2019)053",
    journal = "JHEP",
    volume = "12",
    pages = "053",
    year = "2019"
}

@article{Bennett:2017kga,
    author = "Bennett, Ed and others",
    title = "{Sp(4) gauge theory on the lattice: towards SU(4)/Sp(4) composite Higgs (and beyond)}",
    eprint = "1712.04220",
    archivePrefix = "arXiv",
    primaryClass = "hep-lat",
    reportNumber = "PNUTP-17-A07",
    doi = "10.1007/JHEP03(2018)185",
    journal = "JHEP",
    volume = "03",
    pages = "185",
    year = "2018"
}

@article{	  Kaplan:1983fs,
  author	= "Kaplan, David B. and Georgi, Howard",
  title		= "{SU(2) x U(1) Breaking by Vacuum Misalignment}",
  reportnumber	= "HUTP-83/A069",
  doi		= "10.1016/0370-2693(84)91177-8",
  journal	= "Phys. Lett. B",
  volume	= "136",
  pages		= "183--186",
  year		= "1984"
}

@article{	  Kaplan:1991dc,
  author	= "Kaplan, David B.",
  title		= "{Flavor at SSC energies: A New mechanism for dynamically
		  generated fermion masses}",
  reportnumber	= "UCSD-PTH-91-04",
  doi		= "10.1016/S0550-3213(05)80021-5",
  journal	= "Nucl. Phys. B",
  volume	= "365",
  pages		= "259--278",
  year		= "1991"
}

@article{	  Georgi:1984af,
  author	= "Georgi, Howard and Kaplan, David B.",
  title		= "{Composite Higgs and Custodial SU(2)}",
  reportnumber	= "HUTP-84/A034",
  doi		= "10.1016/0370-2693(84)90341-1",
  journal	= "Phys. Lett. B",
  volume	= "145",
  pages		= "216--220",
  year		= "1984"
}

@article{	  Dugan:1984hq,
  author	= "Dugan, Michael J. and others",
  title		= "{Anatomy of a Composite Higgs Model}",
  reportnumber	= "HUTP-84/A073",
  doi		= "10.1016/0550-3213(85)90221-4",
  journal	= "Nucl. Phys. B",
  volume	= "254",
  pages		= "299--326",
  year		= "1985"
}

@article{Belyaev:2015hgo,
    author = "Belyaev, Alexander and others",
    title = "{Singlets in composite Higgs models in light of the LHC 750 GeV diphoton excess}",
    eprint = "1512.07242",
    archivePrefix = "arXiv",
    primaryClass = "hep-ph",
    doi = "10.1103/PhysRevD.94.015004",
    journal = "Phys. Rev. D",
    volume = "94",
    number = "1",
    pages = "015004",
    year = "2016"
}

@article{DeGrand:2016pgq,
    author = "DeGrand, Thomas and others",
    title = "{One-loop Chiral Perturbation Theory with two fermion representations}",
    eprint = "1605.07738",
    archivePrefix = "arXiv",
    primaryClass = "hep-ph",
    doi = "10.1103/PhysRevD.94.025020",
    journal = "Phys. Rev. D",
    volume = "94",
    number = "2",
    pages = "025020",
    year = "2016"
}

@book{Akaike:1998zah,
    author = "Akaike, Hirotogu",
    title = "{Information Theory and an Extension of the Maximum Likelihood Principle}",
    doi = "10.1007/978-1-4612-1694-0\_15",
    publisher = "Springer",
    address = "New York",
    year = "1998"
}

@article{Jay:2020jkz,
    author = "Jay, William I. and Neil, Ethan T.",
    title = "{Bayesian model averaging for analysis of lattice field theory results}",
    eprint = "2008.01069",
    archivePrefix = "arXiv",
    primaryClass = "stat.ME",
    reportNumber = "FERMILAB-PUB-20-374-T",
    doi = "10.1103/PhysRevD.103.114502",
    journal = "Phys. Rev. D",
    volume = "103",
    pages = "114502",
    year = "2021"
}

@article{Urrea-Nino:2025ztj,
    author = {Urrea-Ni{\~n}o, Juan Andr{\'e}s and others},
    title = "{Flavor mixing in charmonium and light mesons with optimal distillation profiles}",
    eprint = "2502.04977",
    archivePrefix = "arXiv",
    primaryClass = "hep-lat",
    reportNumber = "WUB/25-02, CERN-TH-2025-026",
    doi = "10.22323/1.466.0080",
    journal = "PoS",
    volume = "LATTICE2024",
    pages = "080",
    year = "2025"
}

@article{Ottnad:2025zxq,
    author = "Ottnad, Konstantin and others",
    title = "{$\eta $, $\eta ^\prime $ mesons from lattice QCD in fully physical conditions}",
    eprint = "2503.09895",
    archivePrefix = "arXiv",
    primaryClass = "hep-lat",
    reportNumber = "MITP-25-021, CERN-TH-2025-052",
    doi = "10.1140/epja/s10050-025-01635-0",
    journal = "Eur. Phys. J. A",
    volume = "61",
    number = "7",
    pages = "169",
    year = "2025",
    note = "[Erratum: Eur.Phys.J.A 61, 227 (2025)]"
}

@article{Bennett:2024cqv,
    author = "Bennett, Ed and others",
    title = "{Meson spectroscopy from spectral densities in lattice gauge theories}",
    eprint = "2405.01388",
    archivePrefix = "arXiv",
    primaryClass = "hep-lat",
    reportNumber = "CTPU-PTC-24-11, PNUTP-24/A02",
    doi = "10.1103/PhysRevD.110.074509",
    journal = "Phys. Rev. D",
    volume = "110",
    number = "7",
    pages = "074509",
    year = "2024"
}

@article{Gui:2026adh,
    author = "Gui, Long-Cheng and others",
    title = "{Scalar glueball-$s\bar{s}$ mixing in one flavor lattice QCD}",
    eprint = "2607.17294",
    archivePrefix = "arXiv",
    primaryClass = "hep-lat",
    doi = "10.48550/arXiv.2607.1729",
    month = "7",
    year = "2026"
}

@misc{HiRepSpN,
	author = {},
	title = {{G}it{H}ub - sa2c/{H}i{R}ep: {H}i{R}ep repository},
	howpublished = {\href{https://github.com/sa2c/HiRep}{github.com/sa2c/HiRep}},
	year = {},
}

@misc{HiRepSUN,
	author = {},
	title = {{G}it{H}ub - claudiopica/{H}i{R}ep: {H}i{R}ep repository},
	howpublished = {\href{https://github.com/claudiopica/HiRep}{github.com/claudiopica/HiRep}},
	year = {},
}

@article{UKQCD:2006singlet0pp,
    author        = "Hart, A. and others",
    collaboration = "UKQCD",
    journal       = "Phys. Rev. D",
    volume        = "74",
    pages         = "114504",
    year          = "2006",
    doi           = "10.1103/PhysRevD.74.114504",
    eprint        = "hep-lat/0608026",
    archivePrefix = "arXiv",
    primaryClass  = "hep-lat"
}

@inproceedings{Brett:2019scalarglueball,
    author        = "Brett, Ruair{\'\i} and others",
    journal       = "AIP Conf. Proc.",
    doi           = "10.1063/5.0008566",
    eprint        = "1909.07306",
    archivePrefix = "arXiv",
    primaryClass  = "hep-lat"
}

@article{Michael:2013gka,
    author = "Michael, Chris and others",
    collaboration = "ETM",
    title = "{$\eta$ and $\eta^\prime$ mixing from Lattice QCD}",
    eprint = "1310.1207",
    archivePrefix = "arXiv",
    primaryClass = "hep-lat",
    doi = "10.1103/PhysRevLett.111.181602",
    journal = "Phys. Rev. Lett.",
    volume = "111",
    number = "18",
    pages = "181602",
    year = "2013"
}

@article{juan_mixing_swave,
  title = {S-wave flavor-singlet meson mixing in QCD with light and charm quarks},
  author = {Urrea-Ni{\~n}o, Juan Andr{\'e}s and H{\"o}llwieser, Roman and Knechtli, Francesco and Korzec, Tomasz and Finkenrath, Jacob and Peardon, Michael},
  journal = {Phys. Rev. D},
  volume = {112},
  issue = {7},
  pages = {074502},
  numpages = {14},
  year = {2025},
  month = {Oct},
  publisher = {American Physical Society},
  doi = {10.1103/tcbk-x4f1},
  url = {https://link.aps.org/doi/10.1103/tcbk-x4f1}
}

@article{juan_mixing_charmonium_glueball,
  title = {Charmonium-glueball spectroscopy with improved hadron creation operators},
  author = {Urrea-Ni{\~n}o, Juan Andr{\'e}s and Knechtli, Francesco and Korzec, Tomasz and Peardon, Michael},
  journal = {Phys. Rev. D},
  volume = {114},
  issue = {1},
  pages = {014509},
  numpages = {17},
  year = {2026},
  month = {Jul},
  publisher = {American Physical Society},
  doi = {10.1103/rs22-dx23},
  url = {https://link.aps.org/doi/10.1103/rs22-dx23}
}

\end{document}